\documentclass[12pt]{article}
\usepackage{amssymb,amsmath}
\usepackage[dvips]{graphicx}

\newcommand{\nn}{\nonumber}
\newcommand{\nin}{\noindent}
\def\Pm{\hbox to 6pt {$I\!\!\!\!P~~~~$}}

\begin{document}
\title{\bf On the high-energy Elastic Scattering of hadrons at large t}
\author{ {\sc O.V. Kancheli}\thanks{E-mail: kancheli@itep.ru} \\
     {\it  Institute of Theoretical and Experimental Physics, }  \\
     {\it  B. Cheremushinskaya 25, 117259 Moscow, Russia. } }
\date{}
\maketitle
\begin{abstract}
The main contribution to hard elastic scattering comes from
components of wave functions of colliding hadrons that contain
minimum number of partons. We discuss this mechanism in regge and
parton approaches and estimate the probabilities that colliding
hadrons are in such bare states. The behavior of cross-sections in
this regime at various energies can give nontrivial information on
high energy dynamics.
\end{abstract}

\newpage
\setcounter{footnote}{0}
\section*{\bf 1. Introduction.}

The elastic scattering at high energies is reasonably well
described by the regge approach.  At small $t$ it has the
diffractive nature and is connected with the pomeron exchange. For
the supercritical pomeron \Pm ~~(such structure of \Pm~ is
suggested by experiments and by the QCD calculations) the mean
number of ~\Pm~ exchanges increases with energy. And gradually, at
asymptotic energies,  the power grow of amplitudes $A \sim \exp
(\Delta y)$ changes, as usually expected, to the Froissart type
behavior $A \sim y^2$~, where  $y=\log s/m^2$ is the full
rapidity. The width of diffraction peak $\delta t \sim 1/R^2(y)$
at not very high $y$ is defined by the regge radii of colliding
objects $R^2(y) \simeq 2 R_0^2 + \alpha' y$. The diffractive
contributions decrease fast with $k_{\perp}^2 = -t$ -
approximately as ~$\exp [- k_{\perp}^2 R^2(y) ]$.

The character of interaction changes at subsequent grow of
$k_{\perp}$ - from diffractive to the direct parton exchange.  For
the experimentally investigated energies this transition takes
place at $|t| \simeq 2 \div 5 ~GeV^2$. In regge approach this
transition zone is smeared, because, at first, with grow of
$k_{\perp}$, the mean number of \Pm~ exchanges also grows $\sim
k_{\perp}$, and the elastic amplitude can be approximately
represented by the Orear type expression $\sim \exp(-k_{\perp}
f(y))$. In this case the large transferred momentum $k_{\perp}$ is
distributed approximately uniformly among the exchanged pomerons,
and this is connected with the approximate linearity of regge
trajectories at small $t$.

For even larger $k_{\perp}$ this uniform distribution changes, and
the large transferred momentum  $k_{\perp}$ is concentrated mainly
on a single \Pm~ line. This is primarily related with the
nonlinearity of the ~\Pm~ trajectory and its hard satellites in
QCD, which at large $k_{\perp}$ must approach to the fixed values
as ~$\alpha_P (k_{\perp}) \simeq 1 + c/\log(k_{\perp}^2)$~
\footnote{Note that, as can be seen from a comparison with the
experimental data, the regge trajectories look linear up to rather
large $-t \simeq 2\div 5 GeV^2$, and only after that they can move
to the bare values. The details of this phenomenon are not so well
investigated, and it is, probably, related to the large
nonperturbative contributions, essential up to sufficiently small
distances $\sim 0.1 \div 0.2 ~fm$}.
 In this case the t-behavior of amplitude is defined by the pomeron
vertices, and one can expect the power behavior of full regge
amplitudes $A \sim 1/k_{\perp}^{2\nu}$, where $\nu$ is determined
by the minimal number of exchanges needed to scatter all  parton
components of hadron on the same angle. In this mechanism the main
contribution to the scattering amplitude comes from components of
wave functions of the colliding hadrons with minimal possible
number of partons.
 The picture closely corresponds to the quark counting rules
\cite{qcr} and its development \cite{SatSter}-\cite{land},~  and
the corresponding behavior of cross-sections is approximately seen
in experiments.

Such high $k_{\perp}$ scattering mechanism contains two main
ingredients. One is the amplitude of hard scattering of valent
(bare) constituents. It is defined by the perturbative QCD, and
can be estimated even on the dimensional grounds. The other is the
amplitude to find hadron in such a ``bare'' state, and it is
defined mainly by the nonperturbative physics. What one first of
all needs is the dependence of this amplitude on the energy of
hadrons and the transverse resolution. This is what we consider
below in regge and parton approaches.

\section*{\bf 2. ~~Cross-sections  }
\vspace{1pt}

Consider the contribution to the elastic scattering amplitude
$A(s,t)$ from  rather general reggeon diagrams of Fig.1 ~at high
$|t| \gg 1/R^2$, but such that $|t| \ll s$. Because, as we have
assumed, the \Pm ~trajectories are nonlinear, and freeze at high
$-t$, the \Pm ~exchange amplitude $v \sim i g^2(t)s^{\alpha(t)}$
behaves at high $-t$  in power-like manner
\footnote{ The structure of pomeron in QCD can be rather
complicated and \Pm ~probably consists of sequence of regge poles
$\Pm_n$ with intercepts $\simeq 1 + c/n$, and small $t$ slope
$\sim 1/n^2$, and such that their internal virtuality grows like
$q_{\bot}^2 \sim \exp n$. At high $-t$ all these satellite $\Pm_n$
accumulate at their bare values like $\alpha_{nP} (k_{\perp})
\simeq 1 + c/( n + c_1 \log(k_{\perp}^2/\Lambda_c^2))$. ~At high
momentum transfer the contribution of satellites $\Pm_n$ with high
$n$ in the amplitude can be even dominant - so that the essential
$n \sim  \log (-t)$.~ But here we will not take into account these
details because all $\Pm_n$ trajectories move at high $-t$ to the
same bare value, and the structure of exchange is defined by the
minimal perturbative graph. }.
\begin{figure}[h]
\begin{center}
\includegraphics[scale=0.55, keepaspectratio=true]{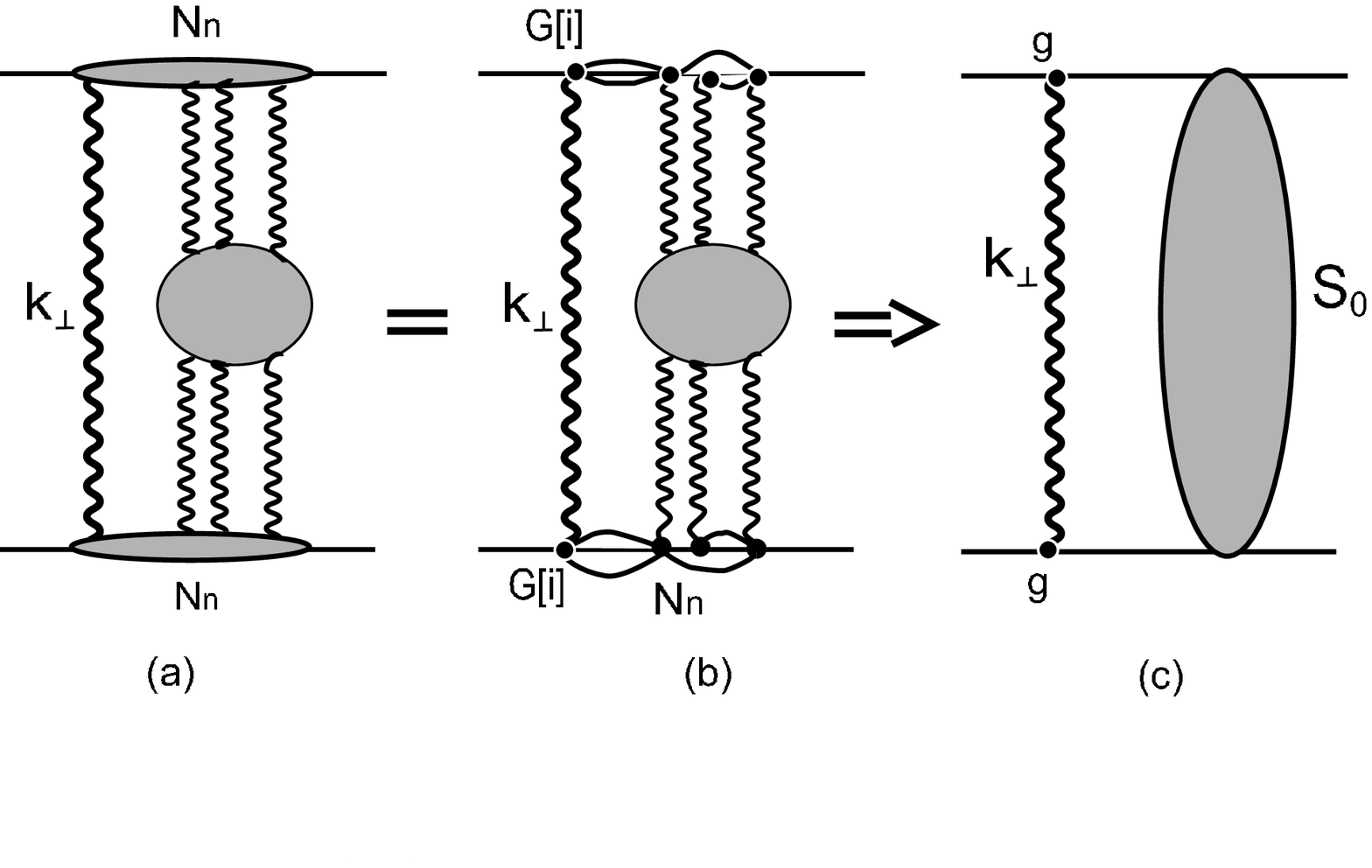}
\parbox{12cm}{    {\bf Fig.1~:}~~
 {\bf (a)} ~General reggeon diagram for elastic~ multi-\Pm ~apmlitude.
 The \Pm ~~line with high $k_{\perp}$ is selected.
 ~~~{\bf (b)} The decomposition of the multipomeron vertexes $N_n$
   into jets.
 ~~~{\bf (c)} Diagram essential at large $k_{\perp}$.
}
\label{Fig1}
\end{center}
\end{figure}
As a result in this case the main contribution to $A$ comes from
diagrams Fig.1a  ~in which almost all transverse momenta is
transferred trough a single \Pm ~line. The full diagram for $A$
can be symbolically represented as in Fig.1c with the contribution
\begin{eqnarray}\label{ampl}
  A(s,t=-k_{\perp}^2) ~=~
 \int d^2 q_{1\perp}  S(s,q_{\perp}^2)
v(s,(k + q)_{\perp}^2)  ~\simeq~~~~~~~~~~~~~~~~~~~~~~~   \nn \\
\simeq~   v(s,k_{\perp}^2) ~\int d^2 q_{1\perp} S(s,q_{\perp}^2)
   ~=~ v(s,k_{\perp}^2) \tilde{S}(s,0)~~,~~~~~~~~~~~
\end{eqnarray}
where we separated  integration over the \Pm ~line  with high
momentum transfer. Here $\tilde{S}(s,b)$ is the elastic S-matrix
in impact parameter plane $b$, and in~(\ref{ampl}) it enters at
zero impact parameter, reflecting in particular that at high
momentum transfer the small $b$ are essential. This finally gives
\begin{equation}\label{equat}
  \frac{d \sigma}{d t} ~\simeq~
         |S_0(s)|^2 ~\frac{d \hat{\sigma}}{d t} ~~,~~~~~~~
           S_0(s)~\equiv~  \tilde{S}(s,b=0)~~,
\end{equation}
where  $d \hat{\sigma}(s,t)/d t$ is the ``bare'' cross-section,
connected  with hard exchange.
 The quantity $\tilde{S}(s,b=0)$, entering (\ref{ampl}),
represents the contribution from the interaction in the initial
and final state, and the $\tilde{S}(s,b)$ is equal to the
amplitude that one of colliding particle moves trough another
without interaction at given $s$ and impact parameter $b$.~ One
factor $|S_0(s)|$ in (\ref{equat}) can be interpreted as a
probability to find the fast particles in such a ``bare'' state
that they almost do not interact with a target. Another factor
$|S_0(s)|$  reflects that scattered particles in the final state
are also in such a bare state. In the lab. frame of one of
colliding particles the $|S_0(s)|$ gives the probability that
another particle is in the state without a soft parton cloud
\footnote{ In logarithmic theories like QCD one should (in
principal) include in a parton cloud also all partons with $q_i^2
\leq k^2_{\perp}$, which can be separately exited in such a hard
process. But at not too high energies there are low partons with
high $k_{\bot}$. It is seen also even at the LHC energies  where
the mean $k^2_{\perp}$ almost do not grow.}.

The essential point which has to be taken into account when going
from diagrams Fig.1a,b to Fig1.c , is connected with the structure
of multipomeron vertices $N_n$.  In the general contribution of
Fig.1b the multiparticle diffractive jet goes out from the hard
\Pm ~vertices $\emph{G}[i]$, and then in resulting expressions
(\ref{ampl}, \ref{dsdt}) one should sum over the states [i] of
such jets.
  But the particles produced in such a hard diffraction would have
large relative transverse momenta of order $k_{\perp}$. Particles
lines with these high $k_{\perp}$ enter neighboring vertices and
will lead to an additional smallness $O(1/k_{\perp}^2)$ of
corresponding contribution to the amplitude. Alternatively, this
jet can be especially aligned in order to compensate the large
transferred momentum to other vertices -- the corresponding small
probability results in the same small factor in multiparticle jet
contribution. Therefore at high $k_{\perp}$ in the states outgoing
from the hard vertices $g$ in Fig.1c only one particle state
survives. Such an answer looks natural because in this case the
minimal number of constituents pass trough the hard vertex.

In the factorized form (\ref{equat}) the quantity $S_0^2$
represents the contribution of large transverse distances and $d
\hat{\sigma} /dk_{\perp}^2$ of the small ones. At very high
$k_{\bot}$ and $s$ the hard cross-section or the $S_0^2$ can
additionally contain the Sudakov like suppression factor depending
on the scale corresponding to the border between small and large
distances
\footnote{ In QCD the scale separating nonperturbative and the
perturbative regions is probably located at rather high
virtuality~  $\kappa \simeq 2 \div 3~GeV$.  In this case the
perturbative Sudakov exponent contains factor $\alpha_s (\kappa)
\log ( k_{\bot}/\kappa ) $ , which is not sizeable.}.

The cross-section of the hard \Pm-exchange  $d \hat{\sigma}(s,t)/d
t \sim g_1^2(t)g_2^2(t)$ is mainly determined by the behavior of
\Pm~-vertexes $g_j(t)$, because at large $-t$ the purely regge
part
\begin{equation}\label{regge}
s^{\alpha(t)-1} \sim \exp ( ~c~\Delta_P ~\log s /\log (-t) ~)
~~,~~~~c \sim 1
\end{equation}
grows very slowly with $s$. ~In perturbative QCD the hard part of
\Pm ~can be approximately represented as the BFKL ladder with mean
interval between rapidities of emitted gluons of order of $\sim
1/\alpha_s(t)$. Such a BFKL type  \Pm ~exchange is almost of the
same form as the direct 2n gluon exchange; the $n >1$
contributions can arise because the BFKL pomeron contains also the
multigluon t-channel contributions due to gluon reggeization
\footnote{At high $-t$ not only \Pm ~but also the other multigluon
QCD reggeons  can equally contribute to the hard exchange in the
same way as BFKL pomeron ~\Pm~ and odderon. Probably all such
n-gluon trajectories $\alpha_n(t)$ at large $-t$ are approximately
degenerated and accumulate near intercept equal one. The
difference in their contributions comes only from different values
of the corresponding regge vertices.}.
And on this way we come finally to the quark counting type model.

The behavior of $g_i^2(t)$  at large $|t|$ in QCD can be estimated
even quasi-classically, and this approach leads finally to the $t$
dependence~:
\begin{equation}\label{dsdt}
d \hat{\sigma}(s,k_{\perp})/dk_{\perp}^2 \sim
(\alpha_s(k_{\perp}))^{\nu}
     / (k^2_{\perp})^{N}~~~,
\end{equation}
with
$$
 \nu =  n_1 + n_2 + | n_1 - n_2|  ~~,~~~~
        N = \frac{1}{2} \big( 3 ( n_1 + n_2) +| n_1 - n_2|\big) - 1~~,
$$
where $n_1$ and $n_2$ are numbers of valent constituents (fast
quarks) in colliding hadrons. ~In (\ref{dsdt}) we have neglected
the regge factor (\ref{regge}) , which at high $-t$ grows very
slowly with $s$, and have also neglected the possible Sudakov type
suppression factors - they can even approximately compensate one
another. ~~Expression (\ref{dsdt}) leads to the behavior of $\sim
1/t^8$ type for $p p$ and to $\sim 1/|t|^7$ for $\pi p$ cases.

There also can be another transverse configuration of
constituents, in which the scattering index entering
Eq.(\ref{dsdt}) is less then $N$. It corresponds to a scattering
in a state in which the hadron constituents are arranged in the
transverse plane on the line perpendicular to a scattering plane
\footnote{In fact all what is needed and what leads to expression
(\ref{ind}) is the following~: all valent constituents must
scatter at closed angles $\theta\simeq k_{\perp}/k_z$ such that
all relative transverse momenta of these partons be $\lesssim m$.
Also one must fulfil the condition that after scattering all
constituent partons should be located in the same packet of
longitudinal size $\sim 1/k_z$. It leads to the condition that
hadrons predominately scatter in specific initial configurations,
when their partons are arranged on the lines in transverse planes.
These lines must be perpendicular to the scattering plane;  but
the relative separation of partons on lines can be arbitrary. Such
a picture of hard elastic scattering similar to the Landshoff
mechanism \cite{land}}.
In such a case we have
\begin{equation}\label{ind}
N ~\rightarrow~   \hat{N} ~=~ \frac{5}{4}~|n_1 + n_2| +
       \frac{3}{4}~|n_1 -  n_2| -\frac{1}{2}~~,
\end{equation}
which leads to the behavior $\sim 1/|t|^7$ type for $p p$ ~and to
$\sim 1/|t|^{13/2}$ ~for $\pi p$ cases.

Note that the bare hadron cross-section $d \hat{\sigma}/
dk_{\perp}^2$ given by (\ref{dsdt}) can be approximately
interpreted as the cross-section of specific quasi-elastic
process, where in the final state we have two hadrons with high
transverse momenta~ $\simeq \pm k_{\perp}$, and all other
particles have small $k^2_{\perp} \sim  \langle k^2_{\perp}
\rangle$. The particles which have large transverse momenta
$k_{\perp}^2 \gg ~ \langle k_{\perp}^2 \rangle$ and take more a
than half of the total energy originate from hard interactions of
leading partons. All other final particles are soft and come in
the configuration typical for mean inelastic events at the same
energy. These particles are created by the standard hadronisation
of soft parton cloud, which follows after ``removing'' the leading
partons in a hard interaction.

Such events also resemble the final state containing two high
energy jets in special configurations, in which all the jets
energy is concentrated on one fast particle (pion or nucleon). The
probability of jet coming in such an ``empty'' state very likely
contains the same damping factors $d \hat{\sigma}/dk_{\perp}^2$
and $S_0(s)$.

It can be interesting to single out similar events in high-energy
heavy ion interactions. For example, in $A_1 A_2$ collisions those
are the events with two nucleons of high transverse momenta $\pm
k_{\perp}$ and soft rest hadrons with transverse momenta typical
for soft  $A_1 A_2$ collisions. The cross-sections of such a
process is
$$
d \sigma/ dk_{\perp}^2 ~\simeq~  A_1 A_2  \cdot
     ~d \hat{\sigma}/ dk_{\perp}^2~~,
$$
because the nuclei are transparent for the bare nucleons.

\section*{\bf 3.~~   The estimation of  $|S_0|^2$ in parton models}

Thus, to understand the behavior of $d\sigma/dt$ it is essential
to estimate the behavior of $|S_0|$, which gives the probability
that fast hadron is in a bare state, it is it contains only valent
components and has has no additional parton cloud. For all
accessible energies it is, mainly, the soft parton cloud connected
with ~$\Pm$ ~exchange.

The \Pm ~-ladder corresponds, roughly speaking, to a soft
parton(gluon) cascading with some mean step in rapidity
$\delta_y$. Because these cascading steps are almost independent,
the probability that no cascade be generated (it is a state
without additional partons is realized) is of Poisson type $\sim
\exp (- y \tilde{\Delta} )$, where $\tilde{\Delta} = c/\delta_y $
~,~~ $c \sim 1$. It corresponds to the fact that at y-boost the
mean number of low energy partons $n(y)$ is defined by the linear
equation $\partial n(y)/\partial y = \tilde{\Delta} n(y)$
\footnote{Evidently this is so for supercritical \Pm. If
$\alpha_P(0)$ where $< 1$ then the probability find hadron in the
``bare'' state is always finite and dos not decrease with the
hadron energy.}.
~So the crucial quantity is the value of $\tilde{\Delta}$, and we
see that $\tilde{\Delta} \simeq \delta_y^{-1}$ where $\delta_y$ is
the mean step between essential degrees of freedom in the pomeron
ladder. And one can estimate $|S_0(y)| \simeq \exp (- y /\delta_y
)$.

These arguments can be presented in a slightly different way. To
scatter at large $k_{\bot}$ hadrons should have the minimal parton
clouds. It means that colliding particles must fluctuate to a
state with small transverse size $r(y)$. The probability $w$ of
such a fluctuation is $\sim (r(y))^{2(\nu-1)}$, where $\nu$ is the
number of valent constituents. In this case, the mean rapidity
interval between steps in parton ladder would be $\delta_y \sim
1/\alpha_s(r(y))$. From the condition $\delta_y \simeq y$  (no
partons except of valent ones in the whole rapidity interval )  we
find that $r(y) \sim \exp (-c_1 y)$, and we come to the same type
of the exponential dependence $S_0(y) \sim w(y) \sim \exp (-2 c_1
(\nu-1) y)$ as before
\footnote{This type of reasoning can remind the approach
\cite{bpst} to high energy scattering used in dual models embedded
in the $AdS_5$ space.}.

Additional arguments for exponential behavior of probability $S_0
(y)$ come from requirement of boost invariance of hard collision
description in partonic terms. Let us consider this in arbitrary
longitudinal frame, when the colliding particles have rapidities
$y_1$ and  $y_2$. Then the quantity $S_0(y)$, which is the
amplitude that both colliding particles are in a bare state must
have the multiplicative form
$$
       S_0(y)  ~=~  S_0(y_1)~S_0(y_2)~~,
$$
where $y = y_1+y_2$. ~Because at boost $y_1 \rightarrow y_1 + \eta
, ~y_1 \rightarrow y_1 - \eta, y \rightarrow y $,~ the only
functional form of $S_0(y)$ that fulfils this condition in
arbitrary frame is exponential $S_0(y) \sim \exp( c y)$.

So to estimate the behavior of $S_0(y)$ we need to know the mean
number of steps in the pomeron ladder. Two possibilities can be
emphasized, ~and they correspond to a different choice of partons
- the independent degrees of freedom in the Fock wave function of
fast hadron.

In one case we can consider the Pomeron ladder as constructed from
white particles (for example, from $\pi$ and $\rho$ mesons), as it
is usually done in various multiperipheral type models. In this
case the steps (parton) density in rapidity is ~$\simeq 0.5 \rho
\simeq 1$, where $\rho$ is the mean density of produced hadrons at
not too high energies, when the one \Pm ~exchange dominates.

In other case one can identify the soft \Pm ~ladder with the rare
gluon ladder with steps $\tilde{\delta}_y \simeq \Delta_P^{-1}$,
where $\Delta_P \simeq 0.1 \div 0.2$ is the ``experimental'' soft
pomeron intercept
\footnote{Such type of model was proposed \cite{AbraKa} long ago,
and in this case (with large $\delta_y$) one can simply explain
unnaturally small value of various pomeron parameters, such as
$\alpha_{P}',~ \Delta_{Pm},~r_{3P}$,... }.

It seams, that the second possibility is theoretically more
preferable, because the white hadron-partons do not represent the
independent degrees of freedom, and therefore they are not
completely appropriate to be used in the Fock space Hamiltonian.

If hard partons were also essential in the Fock w.f. then at first
sight instead of $\simeq \exp (- y /\delta_y )$  one can expect a
more complicated behavior $|S_0(y)|$. But the hard gluons are
mainly taken into account in the parton w.f. as constituent of
soft partons. It can be seen from the fact that at all acceptable
energies the mean transverse momenta of secondary particles almost
do not grow with energies. We come to the same conclusion from a
different way by remarking that the hard gluon spectra
``measured'' in deep-inelastic reactions can be successfully
described \cite{CapKaid} as coming from renormalization group
rescaling of the soft parton component.

At asymptotic energies the soft parton saturation can become fully
essential and as a result the saturation scale $Q^{(sat)}(y)$ in
transverse momenta can also become large. Then one should take
into account in $|S_0|^2$ also  more hard partons. Note that even
in the in the asymptotical Froissart regime where saturation
dominates the behavior of $|S_0|$ can be estimated in the same way
as the probability that the valent components do not emit any
primary gluons in the corresponding ranges of energy and
transverse momenta ($ q_{\perp}< \langle k_{\perp}\rangle \sim
Q^{(sat)}(y) $). This is quite enough, because all other partons
are emitted by these primaries. This evidently leads to a mean
number of the primary partons
$$
\bar{n}(y,k_{\perp}) \sim \int d\omega/\omega \int^{k_{\perp}}
dq^2_{\perp} \alpha_s(q_{\perp})/ q_{\perp}^2
$$
and to Sudakov type factor for the no-emission probability
$W(y,k_{\perp}) \sim \exp (-\bar{n})$, and, eventually, to the
estimate $|S_0(y,k_{\perp})| \simeq W(y_1,k_{\perp})
W(y_2,k_{\perp})$, which is again boost-invariant.

\section*{\bf 4.~~ The estimation of  $|S_0|^2$ in regge approach}
\vspace{2mm}

The numerical values of $S_0 \equiv |S_0(y,b=0)|$ can be estimated
directly from the experimental data on behavior  of the profile
function $F(y,b)=1-S(y,b)$ at $b=0$, calculated by the Fourier
transformation of $\sqrt{(d\sigma/dt)^{(exper)}}$. For the $pp$
and $p\bar{p}$ scattering it ranges  from values $S_0 \simeq 0.6$
at $\sqrt{s} \sim 50 GeV$ up to values $S_0 \simeq 0.01 \div 0.02$
at $\sqrt{s} \sim 2~TeV$.

In regge models the value of $S_0$ depends crucially on the
relative weights of contributions of multipomeron exchanges to the
elastic amplitude, and, in principle, can be extracted from
``every'' good model descriptions of $d\sigma/dt$ data at low $t$.
But it is essential that in the most popular models one can not
expect the functional behavior $S_0(y) \sim \exp (- y c )$ at
large $y$, which seams very natural in the parton approach.

If $v(y,b)$ is the one \Pm ~contribution to the amplitude, then
the full sum of all pomeron diagrams (neglecting the pomeron
interactions with one another) can be approximately represented as
some function of $v$ in the form
\begin{equation}\label{sm}
S(y,b) ~=~ S[v] ~=~ \sum_{n=0}^{\infty} \frac{\gamma_n}{n!}~(iv)^n
~~,~~~~~~ \gamma_0 = \gamma_1 = 1~,
\end{equation}
where the quantities $\gamma_n > 1$ for $n\geq 2$, and they take
into account the contribution of diffractive jets in multipomeron
vertices $N_n$.

In the simplest eikonal case, when all $\gamma_n =1$  we have $S =
\exp (iv)$ and it leads to $|S_0(y,b)| = \exp (- \verb"Im"
~v(y,b)) \sim \exp (-c_1 \exp(\Delta y))$. Although such an
eikonal-type amplitude (especially if properly adjusted
\footnote{The simplest ``popular'' form of corrections for the
simple eikonal is the quasi-eikonal picture, in which one
explicitly introduce the bare parton sate, or various multichannel
eikonal models.} )
can lead to reasonable description of many data, there will be the
discrepancy with parton picture at very large $y$ in any case.

The purely exponential behavior of $S(y,b)$ can take place only if
the eikonal coefficients $\gamma_n$ grow like $n!$ at large $n$.
In this case we have for
\begin{equation}\label{geo}
|S_0(y,b)| \simeq (c + \verb"Im" ~v(y,b))^{-1} \sim \exp(- \Delta
y)
\end{equation}
at large $y$
\footnote{For a reasonable behavior of series (\ref{sm}) the
coefficients $\gamma_n$ considered as function of $n$ should be
analytic for $Re ~n \geq 0$. The asymptotics of $S[v]$ for $v
\rightarrow \infty$ is defined by most right singularities of
$\gamma_n$ over $n$. To have the asymptotics (\ref{geo}) the
function $\gamma_n$ must have a singularity at $n =-1$. The fairly
interesting case corresponds to $\gamma_n = \Gamma(n+1)$  when
$|S_0(y,b)| \simeq (1 + Im ~v(y,b))^{-1}$  }.

The experimental data  on the behavior of $d\sigma/dt$ at high
$|t|$ are not rich, especially at high $s$. The old data on $p p$
show \cite{land} the universal behavior of $d\sigma/dt \simeq 0.1
t^{-8}$ for $2 \lesssim |t| \lesssim 15$ in the energy range
$\sqrt{s} \simeq 30 \div 60 GeV$. At such energies the $S_0$ is
still rather large, and the decrease of $S_0(s)$ with $s$ can in
fact be compensated by the small growth (\ref{regge}) of
$d\hat{\sigma}/dt$. New data at higher $s$ and $-t$ are needed for
better understanding the behavior of $S_0$.

\section*{\bf 5. Conclusion}

We end with few remarks.

The quantity $|\tilde{S}(y \gg 1,b=0)|$ which represents the
probability that a fast hadron with energy $\simeq (m/2) \exp y$
has no soft parton cloud, is interesting from various aspects.
This quantity can be extracted from the behavior of elastic cross
section $d\sigma/dt$ at relatively low $|t|$. It can be also
extracted from the data on the behavior of elastic cross-sections
at relatively large $|t| > 2 \div 5 GeV^2$ ~(far outside the
diffraction peak). All new data on elastic $d\sigma/dt$ are
therefore very interesting from such a point of view, especially
at maximally large (~LHC~) energies.

The  existing data on $d\sigma/dt$ show that the value of
$|S_0(y,b=0)|$ is not so small, as one can expect at first sight
from the large multiplicity of secondary hadrons at the same
energies. It indicates that the high energy hadron wave function
contains a relatively small number of partons in average. This
also means that most of secondary soft hadrons are created only
after collision (that can be a decay of the nonperturbative QCD
tubes or minijets), and are not directly represented by the
corresponding degrees of freedom in the incoming parton wave
function. In most models of multiperipheral type the opposed
picture is supposed usually

The comparison of the two different theoretical approaches to
estimation of  \nin $|S_0(y,b=0)|$ ~~(based on parton picture and
on multipomeron exchange model) suggests that the pure
eikonal-type unitarization (when $S(y,b) = \exp ( i v )$ ) is not
suitable at very high energies. The most simple form, giving the
same answer for $|S_0|$ as in the parton case, is given by the
expression $S(y,b) =(1 + i c v)^{-1}$, and it corresponds to the
case when the eikonal coefficients grows ($\sim n!$) due to large
contribution of diffractive jets in multipomeron vertices $N_n$.
~~Note that such a grow of multi~-~\Pm ~contributions,  ~in
comparison with the eikonal case, can essentially affect the form
of the tail of multiplicity distribution
\footnote{ In the eikonal case  the the behavior of the tail of
multiplicity distribution is roughly $\sigma_n \sim \sigma_{inel}/
\Gamma (1 + n/ \bar{n})$. ~In the case (\ref{geo}) the tail is
larger: $\sigma_n \sim \sigma_{inel}\exp{(- c n/\bar{n})}$. Note
that experimentally the tail has the form  $\sigma_n \sim \exp{(-
3 n/ \bar{n})}$. Here $ \bar{n}$ is the mean multiplicity. } .

\newpage

\vspace{8mm}
\nin {\bf ACKNOWLEDGMENTS} \\
\nin I thank K.G.~Boreskov for conversations and many essential
comments. Discussion with A.B.~Kaidalov and his remarks where also
very useful.

The financial support of CRDF through the grants RUP2-2961-MO-09
is gratefully   acknowledged. \vspace{4mm}



\begin{thebibliography}{99}


\bibitem{qcr}
S.J.Brodsky and G.R.Farrar Phys.Rev.Lett 31 (1973) 1153~,~~

V.A.Matveev, R.M.Muradyan and A.N.Tavkhelidze, Lett.Nuovo Cim.7
(1973)719

\bibitem{SatSter}
G.Sterman,  arXiv:1008.4122~,~~ M.G.Sotiropoulos and G.Sterman,
hep-ph/9401237~,~~~
   Nuclear Physics {\bf B}419(1994)59~,~~{\bf B}425(1994)489

\bibitem{land}
A.~Donnachie and P.V.~Landshoff,   ~{\em Nucl.Phys.} {\bf B244}
(1984) 322;
  ~{\em Z. Phys.} {\bf C 61} (1994) 139~,~~
   hep-ph/9607377

\bibitem{AbraKa} V.A.Abramovsky and O.V.Kancheli,
  Pisma Zh.Eksp.Teor.Fiz.32:498-501,1980.

\bibitem{CapKaid} A.Capella, E.G.Ferreiro, C.A.Salgado,
     A.B.Kaidalov, ~~hep-ph/0005049, ~hep-ph/0006233

\bibitem{bpst}
J.Polchinski and M.J.Strassler, arXiv:hep-th/0209211~,~

R.C.Brower,~J.Polchinski,~M.J.Strassler,~C.I.Tan,
 JHEP{\bf 0712}, 005, 2007 [arXiv:hep-th/0603115]

\end{thebibliography}
\end{document}